\documentclass[vecphys]{svmult}
\usepackage{makeidx}         % allows index generation
\usepackage[dvips]{epsfig}
\usepackage{graphicx}        % standard LaTeX graphics tool
\usepackage{multicol}        % used for the two-column index
\usepackage[bottom]{footmisc}% places footnotes at page bottom
\makeindex             % used for the subject index
\begin{document}

\title*{Characterizing Stellar Populations in Spiral Disks }
%\titlerrunning{Stellar Populations in disks}
\author{Mercedes Moll\'{a}\inst{1}\and Simon Cantin\inst{2}\and
Carmelle Robert\inst{2}\and Anne Pellerin\inst{3}\and Eduardo
Hardy\inst{4}}
% Use \authorrunning{Moll\'{a} et al.} for an abbreviated version of
\authorrunning{Moll\'{a} et al.}

\institute{CIEMAT, Avda. Complutense 22, Madrid (Spain) 
\texttt{mercedes.molla@ciemat.es}
\and Universit\'{e} Laval and Observatoire du mont M\'{e}gantic, Qu\'ebec (Canada)
\and Space Telescope Science Institute, 3700 San Martin Drive, Baltimore (USA)
\and National Radio Astronomy Observatory, Casilla El Golf 16-10, Santiago (Chile)}

\maketitle

\begin{abstract}
It is now possible to measure detailed spectral indices for stellar
populations in spiral disks.  We propose to interpret these data using
evolutionary synthesis models computed from the Star Formation
Histories obtained from chemical evolutionary models. We find that
this technique is a powerful tool to discriminate between old and
young stellar populations. We show an example of the power of Integral
Field spectroscopy in unveiling the spatial distribution of
populations in a barred galaxy.
\end{abstract}

\section{Introduction}

The information obtained for a spiral galaxy may come from two
sources: the gas or the stars. It is usually obtained from emission or
absorption lines respectively (see \cite{hen99} for a review).  In the
case of spiral galaxies the spectra of the H{\sc ii} regions which
will display emission lines of H$\alpha$, [OIII] etc., provide
information on the chemical abundances in the gas and/or the
\emph{present} Star Formation Rate (SFR).  The atomic and/or molecular
gas densities may also be estimated from observations. These data
represent the present time state of a galaxy.

It is currently possible to perform spectrophotometric observations in
different wavelength bands in order to obtain the radial distribution
of surface brightness, colors, or even of spectral indices that
quantify the absorption lines of the stellar disk.  These observations
measure the added light of all the stellar generations still present
on the disk of the galaxy, thus providing information on the
\emph{average} properties of the disk over the galaxy life time.

The question then is how to use these two types of data, emission and
absorption, to determine the history of a spiral galaxy. What are the
possible evolutionary paths followed by a galaxy in order to reach the
present observed state of its youngest generation while displaying the
observed average properties?

\section{Observing Spiral Disks}
\label{obs}

The measurement of the stellar absorption lines indices of a galaxy is
important since these provide information on the averaged properties
of the various stellar populations and thus represents a powerful tool
in braking the age-abundance degeneracy (see next
section). Historically observations of stellar absorption lines in
spiral disks have been difficult to obtain because the flux level from
the sky is usually very close to that emitted by the stellar
populations under study, especially in the outer disk regions.  The
solution has been to obtain narrow-band images with filters specially
designed for each galaxy \cite{beau97} or to perform Fabry-Perot
interferometry with a Taurus-Tunable filter \cite{ryd05}.  Both
methods are conceptually similar, although the instrumentation is
different. The idea resides in measuring fluxes within wavelength
bands corresponding to specific spectral indices, as close in time as
possible with the observations of the corresponding continuum
bands. The galaxy disk is divided in concentric rings and the flux is
then integrated azimuthally so that the effective signal-to-noise for
the specific absorption line increases. A more modern approach is the
use of a 3D spectroscopic instrument such as OASIS \cite{can05} or
SAURON \cite{fal04} which offer high spatial and spectral
resolution. These instruments allow separation of the gas and young
stars emission line regions from the older phases of star formation.

\section{The Degeneracy Problem : Theoretical Models}
\label{how}

The information from the gas phase, such as density, abundance, actual
star formation rate, is usually analyzed using chemical evolutionary
models.  They describe how the proportion of heavy elements
present in the interstellar medium (ISM) increases, starting from
primordial abundances, when stars evolve and die.  Modern codes solve
numerically the equation system used to describe a scenario based on
initial conditions for the total mass of the region, the existence of
infall or outflow of gas, and the initial mass function (IMF).
Stellar mean-lifetimes and yields, known from stellar evolutionary
tracks are also included.  Finally, a star formation rate is assumed.
These many inputs, which greatly influence the final results of a
model, imply that a large number of parameters must be chosen.  The
{\sl best} model for a galaxy will of course be the one which
reproduces the observational data as closely as possible.  The
resulting star formation history (SFH) and the age-metallicity
relation (AMR) might be taken as the basic relations to describe the
evolution of the galaxy.

Chemical evolutionary models, however, suffer from the well known {\sl
uniqueness} problem: it is usually impossible to describe the
evolution of a galaxy only by knowing the final state or present day
abundances, since more than one evolutionary scenario can be used to
reach the present situation.  A solution is feasible however if the
number of observations is larger than the number of parameters used to
constrain the model.

On the other hand, data related to the stellar phase, such as the
radial distribution of the surface brightness or colors, are usually
analyzed through (evolutionary) synthesis models -- see \cite{del05}
for a recent and updated evolutionary synthesis model --, based on 
single stellar population (SSP) created by an instantaneous burst of
star formation.  These codes compute the spectral energy distribution
(SED), $S_{\lambda}$, colors and spectral absorption indices emitted
by a SSP of metallicity $Z$ and age $\tau$, from the sum of spectra of
all stars created and distributed along a HR diagram, convolved with
an initial mass function. This SED is a characteristic of each SSP of a
given age and metallicity.

These models suffer, in turn, from a {\sl degeneracy} problem: the old
and metal-poor SSP show a similar SED than the young and metal-rich
ones.  A possible solution is to use absorption spectral indices, since
some of them depend on $Z$ while some others, particularly the Balmer
and other hydrogen lines, depend only on $\tau$. Thus, with two
orthogonal indices, it is possible to find the parameters ($\tau$,
$Z$) for a SSP.  This technique was first developed and used to study
globular clusters and elliptical galaxies, for which it was assumed
that the star formation occurred in a short and early burst. It seems
possible from this scheme to find when the burst took place, based on the age of the
stellar population, and also what  the metallicity of the stars is.

\section{Applying Synthesis Models to Spiral Disks}

When the star formation does not occur in a single burst, as it seems
to be the case in spiral galaxies where the star formation is
continuous, the characteristics found using a SSP model represent only
averaged values. These are weighted by the luminosity of each stellar
generation, since the final SED corresponds to the one emitted by the
sum of different SSP's when a convolution with the SFR, $\Psi(t)$, is
done :

\begin{equation}
F_{\lambda}(t)=\int_{0}^{t} S_{\lambda}(\tau,Z)\Psi(t')dt',
\label{Flujo}
\end{equation}
where $\tau=t-t'$.

We must find, e.g. using a least square method, the best superposition
of these SSP's which will fit the data. We must estimate the
proportion of each SSP defined by an age and a metallicity which
better reproduce the observation.  This would give the SFH, $\Psi(t)$,
and AMR (i.e. $Z(t)$).  However, this method, in addition to being
costly in computing-time, may produce some unphysical solutions. Due
to this the SFH is usually assumed, not computed, e.g. an
exponentially decreasing function of time.  This also requires some
hypotheses about the shape and the intensity of the SFR or, to avoid a
bias, the adoption of many functions.  This last option leads to an
excessively large number of models.  A second point, usually
forgotten, is that $S_{\lambda}(\tau,Z)= S_{\lambda}(\tau,Z(t'))$,
that is, the metallicity changes with time since stars continue to
form.  Therefore, it is not clear which $Z$ must be selected at each
time step without knowing how it is evolving in the galaxy. Usually,
only one $Z$ is used for the whole integration.

We suggest here a method to explore spiral galaxies, which are very well
studied objects from the chemical evolutionary point of view. Since a
large number of emission data is available, it is possible to compute and
constrain chemical evolutionary models. The SFR and the AMR obtained
as a result of these models may be used as an input for
Eq.~\ref{Flujo}.  If the SED, colors or absorption indices are well
fitted simultaneously with the present--time data, we then have a good
characterization of the evolutionary history of the galaxy.

We applied our technique to three Virgo galaxies for which the radial
distribution of the spectral indices Fe5270 and Mg2 had been observed
\cite{beau97,mol99}.  We first computed the multiphase chemical
evolutionary models for these galaxies, using a code that was
developed for the Solar Neighbourhood \cite{fer92} and later applied
to the Milky Way Galaxy (MWG) and to other spiral disks (see
\cite{mol05} and references therein).  As we explained in
\cite{mol99}, the radial distribution of the spectral indices were
predicted using Eq.\ref{Flujo} where the SFH and AMR result from
the best chemical evolution model.

In order to study the problems of uniqueness and degeneracy of these
models, we performed 500 different models for one of the three Virgo
galaxies: NGC~4303 \cite{mol02}. From a statistical analysis using a
$\chi^{2}$ method, we determine that only $\sim$20 ( a 4\%) of these
models can reproduce the present day data.  The evolutionary synthesis
code was then applied to these 20 models to calculate the radial
distribution of the Fe5270 and Mg2 spectral indices. Only $\sim$6 (a
1\%) of them were able to fit simultaneously both sets of
observations. Furthermore, these latter models defined a very small
region in the parameter space. This implies that the SFR and AMR
obtained for each radial region of the galaxy can be known with a high
level of confidence. Our technique is therefore quite powerful in estimating
the evolutionary history of a spiral galaxy when both emission and
absorption lines data are available.

\section{The Barred Galaxies: NGC~4900 and NGC~5430}

We have observed two barred galaxies NGC~4900 and NGC~5430 with OASIS,
an optical integral-field spectrograph at the Canada-France-Hawaii
Telescope \cite{can05}.  About 1000 spectra in the wavelength ranges
from 4700 to 5500\AA\ and from 6270 to 7000\AA\ have been collected
with a spatial resolution of 0.80$^{\prime\prime}$ within the central
12$^{\prime\prime} \times 17^{\prime\prime}$ of each galaxy.
\begin{figure}
\includegraphics[width=.5\columnwidth, angle=0]{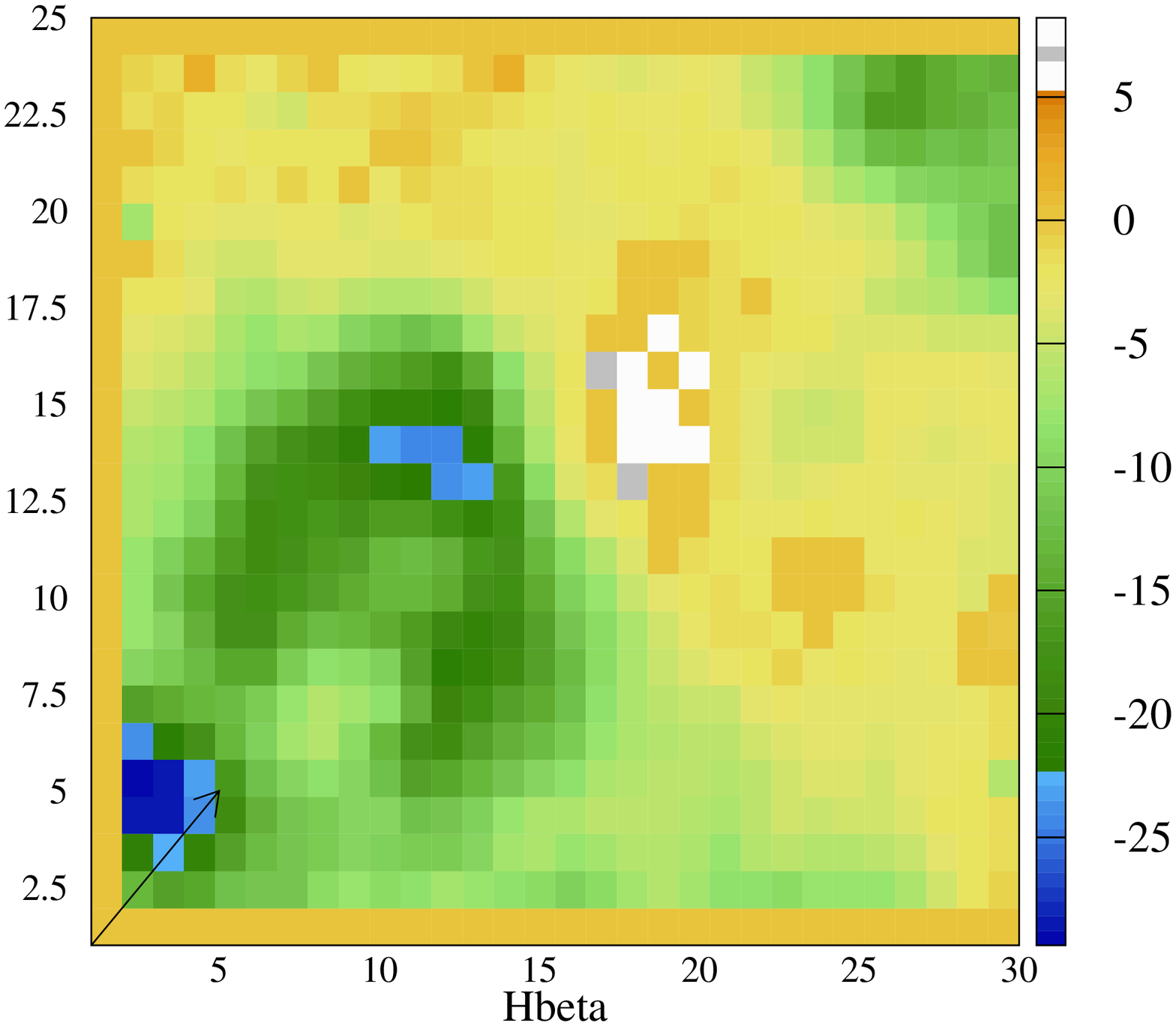} \hfil
\includegraphics[width=.5\columnwidth, angle=0]{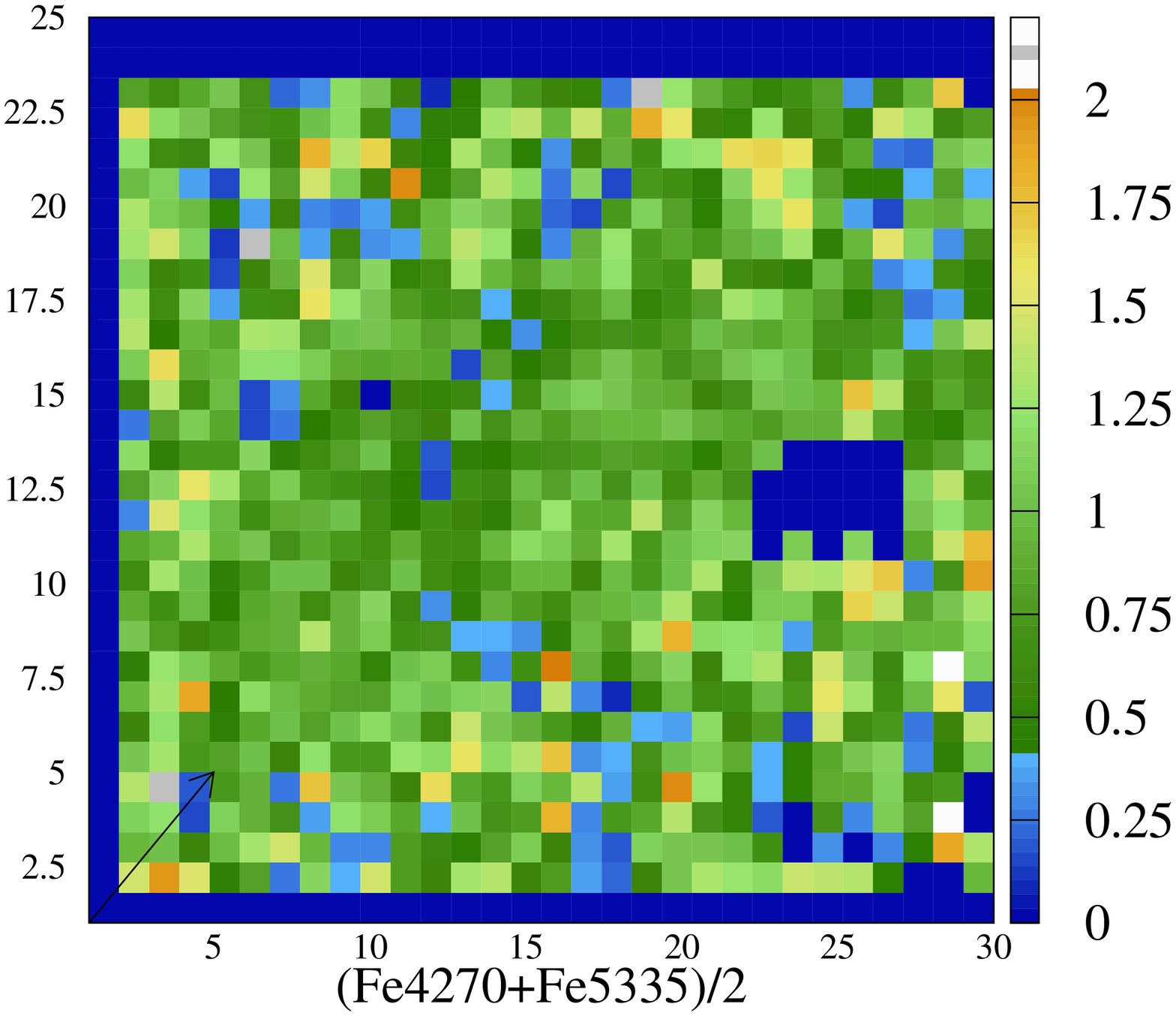}
\caption{Maps of spectral indices H$\beta$ and $<Fe>=(Fe5270+Fe5335)/2$
in units of \AA. One element in $x$ and $y$  covers 100~pc.}
\end{figure}
As shown in Figure~1, these data allow us to create detailed maps for
the different spectral indices showing various morphologies.  When
compared with model predictions, the values for the H$\beta$, $Fe5270$
and $Fe5335$ spectral indices allow us to locate the gas and the
stellar populations.  Thus, in NGC~4900, the youngest stellar
population (as described by H$\beta$, left panel of Fig.1) are located
parallel to the galaxy large scale bar (marked by the arrow), while
the older stars, as described by Fe5270 and Fe5335 spectral indices,
are distributed around the bar.

\section{Conclusions}

\begin{enumerate}
\item Measurements of spectral indices in spiral disks are now
possible \cite{beau97,ryd05}, either through narrow-band imaging,
mostly with a Fabry-Perot interferometer and a Taurus Tunable filter,
or using 3D spectroscopy.

\item The use of chemical evolutionary tracks and evolutionary
synthesis models to interpret these data results in a powerful tool to
determine with precision the evolutionary history of spiral galaxies.

\item The properties of spatially-resolved spiral regions,
as the central region of barred galaxies, observed using 3D spectroscopy,
may be related to other processes, e.g. to a bar which
may produce new star forming bursts. These effects may be quantify
using our combined evolutionary models.

\end{enumerate}

% BibTeX users please use
%\bibliographystyle{cite.sty}
%\bibliography{proc.bib}
%
% Non-BibTeX users please follow the syntax
% the syntax of "referenc.tex" for your own citations
%%%%%%%%%%%%%%%%%%%%%%%%%%%%%%%%%%%%%%%%%%%%%%%%%%%%%%%%%%%%%%%%%%%%%%  }

%%%%%%%%%%%%%%%%%%%%%%%%%%%%%%%%%%%%%%%%%%%%%%%%%%%%%%%%%%%%%%%%%%%%%%

\printindex
\end{document}